\documentclass[article]{aastex6}

\begin{document}

\title{A Double-lined M-dwarf Eclipsing Binary from CSS $\times$ SDSS}

\author{Chien-Hsiu Lee\altaffilmark{1}}
\affil{Subaru Telescope, NAOJ, 650 N Aohoku Pl, Hilo, HI 96720, USA}

\altaffiltext{1}{leech@naoj.org}

\begin{abstract}

  Eclipsing binaries offer a unique opportunity to determine basic stellar properties. With the advent of wide-field camera and all-sky time-domain surveys, thousands of eclipsing binaries have been charted via light curve classification, yet their fundamental properties remain unexplored, mainly due to the extensive efforts needed for spectroscopic follow-ups. In this paper we present the discovery of a short period (P=0.313 days) double-lined M-dwarf eclipsing binary, CSSJ114804.3+255132/SDSSJ114804.35+255132.6, by cross-matching binary light curves from Catalina Sky Surveys and spectroscopically classified M dwarfs from Sloan Digital Sky Survey. We obtain follow-up spectra using Gemini telescope, enabling us to determine the mass, radius, and temperature of the primary and secondary component to be M$_1$ = 0.47$\pm$0.03(statistic)$\pm$0.03(systematic) M$_\odot$, M$_2$ = 0.46$\pm$0.03(statistic)$\pm$0.03(systematic) M$_\odot$, R$_1$ = 0.52$\pm$0.08(statistic)$\pm$0.07(systematic) R$_\odot$, R$_2$ = 0.60$\pm$0.08(statistic)$\pm$0.08(systematic) R$_\odot$, T$_1$ = 3560$\pm$100 K, and T$_2$ = 3040$\pm$100 K, respectively. The systematic error was estimated using the difference between eccentric and non-eccentric fit. Our analysis also indicates that there is definitively 3rd-light contamination (66\%) in the CSS photometry. The secondary star seems inflated, probably due to tidal locking of the close secondary companion, which is common for very short period binary systems. Future spectroscopic observations with high resolution will narrow down the uncertainties of stellar parameters for both component, rendering this system as a benchmark in studying fundamental properties of M dwarfs.
  
\end{abstract}


\section{Introduction}
M dwarfs reside on the cool, low mass end of the main sequence and are 
the most abundant stars in the Milky Way and prevail in
the solar neighborhood. However, due to their faintness in the optical bands, 
they are difficult to discover and characterize. Thanks to the advent of 
wide-field cameras/spectrographs and the on-going large sky surveys,
such as Sloan Digital Sky Survey \citep{2000AJ....120.1579Y}, tens of 
thousands of M dwarfs have been spectroscopically confirmed \citep{2011AJ....141...97W}, yet only a handful of them have basic stellar parameters -- i.e. mass, radius, and effective temperature -- well determined at 3\% level or better \citep{2010A&ARv..18...67T}.

Advances in the M dwarf spectroscopic observations also post 
challenges to theoretical modeling. For instance, discrepancies between 
theoretical modeling and observations of basic parameters of 
M dwarfs have been reported. The models underestimate the size and 
overestimate the temperature at a given M dwarf 
mass, whether it is single, isolated M dwarf or a binary \citep{2012ApJ...757..112B}. Such discrepancies are especially the cases for very low mass star
(VLMS, M$<$0.3M$_\odot$), as shown by \cite{2015MNRAS.451.2263Z}. This is because the stars become completely convective and therefore difficult to model.

In addition, it has been observed the M dwarfs exhibit strong magnetic activity, 
which might be linked to their inflated size \citep{2013ApJ...779..183F,2014ApJ...789...53F}. Estimating fundamental properties of 
low mass stars has drawn increasing interest, in particular more exoplanets 
discovered by Kepler mission are hosted by low mass stars \citep{2013ApJ...767...95D} 
For transiting exoplanets, we rely on precise and accurate determination of the host star 
parameters so as to infer the radius of the exoplanets from transit depth. 
In this regard, improving theoretical modeling of low mass stars, or establishing an
empirical mass-radius and mass-temperature relation for M dwarfs \citep[e.g.][]{2000A&A...364..217D,2016AJ....152..141B,2015ApJ...804...64M,2015ApJ...800...85N}, are highly demanded. 

\begin{figure*}
  \includegraphics[scale=0.7]{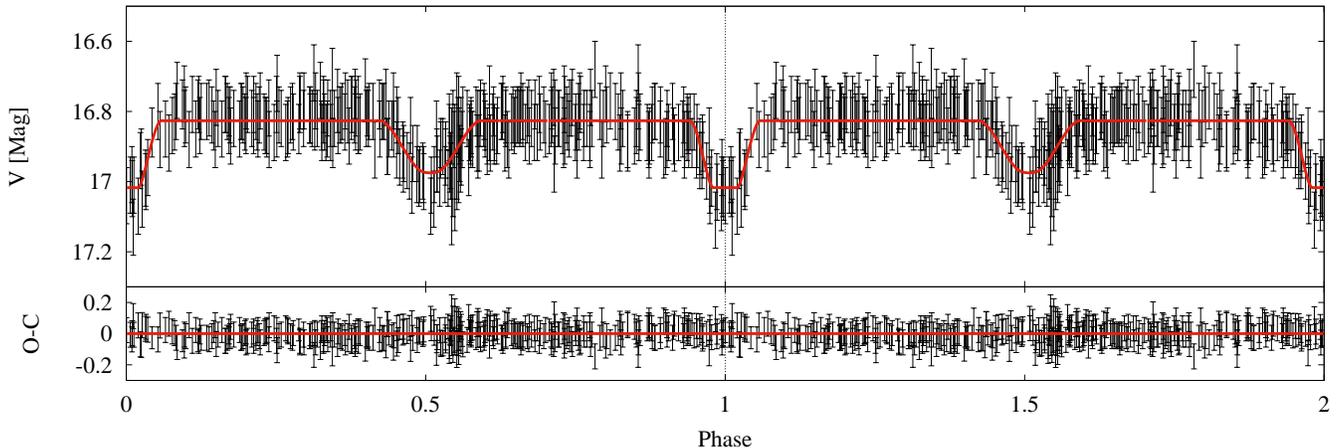}
  \caption{Catalina Sky Survey light curve of the M-dwarf binary system. For illustrational purpose, we plot two full phases to better show the eclipses, with duplicate data shown in phase=1 to phase=2. The vertical dashed line at phase=1 marks the repeated phases. Upper panel: the CSS photometry are marked in black,
    while the best-fit DEBiL model is marked in red. Lower panel: residuals of the best-fit DEBiL model.}
  \label{fig.lc}
\end{figure*}

Eclipsing binaries provide us unique opportunities to independently and accurately measure the mass,
radius, and temperature of individual stars. The
photometric measurements can reveal the information on the inclination angle, orbital period, eccentricity, mass
ratio, and radius in terms of the orbital distance. On the other hand, the spectroscopic
observations enable us to derive the mass and temperature of individual stars, as well as their
orbital distance. Note the normal degeneracy of extracting both temperature and the gravity
(log$g$) from spectra of a single star is not an issue for eclipsing binaries, because we can
determine the mass and radius independently from the radial velocity and light curve alone, hence
unambiguously narrowing down the temperature of the binaries.
Using M dwarfs in eclipsing binaries, \cite{2015MNRAS.451.2263Z}
  presented a summary of 9 M dwarfs, with mass ranging from 0.14 - 0.28 M$_\odot$,
  indicating inflated radius and cooler temperature when compared to stellar models. To further resolve the discrepancy between the stellar models and observations, or to establish an empirical mass-radius and mass-temperature relation for
  VLMS, it is essential to obtain stellar parameters of a larger sample of M dwarfs, preferably from studies of eclipsing binaries.

  In addition, short-period binaries (as presented here) are likely to have inflated radius. Though they are
    not typically used in empirical calibrations, they are important for understanding the interior structures of M dwarfs
  and their inflation mechanisms.

This work reports the discovery of a double-lined M dwarf
eclipsing binary system by joining the forces of Catalina Sky Surveys and Sloan Digital Sky Survey,
as well as dedicated follow-ups using Gemini telescope.
This paper is organized as follows: in \textsection \ref{sec.data} we describe the photometric and
spectroscopic observation in hand. We present our analysis in 
\textsection \ref{sec.ana}, followed by a summary and prospects in \textsection \ref{sec.sum}.

\section{Data}
\label{sec.data}

\subsection{Known M dwarfs from SDSS}
Due to their faintness in the optical bands, previous searches of M dwarfs were limited to infrared imaging studies
\citep[see e.g.][]{2011AJ....142..138L,2013PASP..125..809T}.
Thanks to the advances in wide-field cameras, multi-object spectrographs, and all sky surveys with medium-size telescopes,
it is possible to obtain spectra to confirm M dwarfs in quantity and quality. \cite{2011AJ....141...97W} have taken the
advantage of Sloan Digital Sky Survey \citep[hereafter SDSS][]{2000AJ....120.1579Y}, and present a sample of $\sim$ 70,000
spectroscopically confirmed and classified M dwarfs. Their sample were drawn from $\sim$ 120,000 M dwarf candidates pre-classified by
Hammer spectral typing facility \citep{2007AJ....134.2398C}; After visual inspection, they excluded low S/N spectra and
removed contaminations from extra-galactic interlopers, K and L dwarfs, as well as white dwarf - M dwarf pairs, and retain
70,841, with spectral type as precise as $\pm$1 subtype. This is by far the largest and purest spectroscopic M dwarf catalog,
providing us a firm basis to cross-matching with eclipsing binaries charted by other time-domain surveys.

To search for candidate VLMS in eclipsing binary systems, we then cross-matched the M dwarf catalogue from \cite{2011AJ....141...97W} to the largest all-sky survey eclipsing binary catalogues from the Catalina Sky Survey (see the next section). We specifically searched for dwarfs with spectral type of M3V or later, corresponding to an average stellar mass $<$0.3M$_\odot$ according to \cite{1996ApJ...461L..51B}. We note that the corresponding mass of a given spectral type in \cite{1996ApJ...461L..51B} has a large range, and we only used the average value of that mass range. Thus, it is likely that we selected M3V type stars, but they are still heavier than 0.3 solar mass. As CSS is brightness-limited, the intrinsically brighter early M dwarfs are favored. Thus, it is more likely that we will select stars at the high mass end of our limits. In the end, we found only two M3V dwarfs in eclipsing systems: CSS114804.3+244132/SDSSJ114804.35+255132.6 (presented here) and CSSJ1156-0207 (still under investigation).

\subsection{Time series photometry} \label{sec.lc}
Since 2004, the Catalina Sky Survey employed three small telescopes to survey sky regions between -75 $<$ Dec $<$ 70 deg. Two of the three telescopes patrol the northern sky from Arizona, USA: the Catalina Schmidt Telescope (0.7m, equipped with a 8.2 deg$^2$ FOV camera) the Mount Lemmon Telescope (1.5m, equipped with a 1 deg$^2$ FOV camera). In addition, the Siding Spring Telescope (0.5m, equipped with 4.2 deg$^2$ FOV) surveys the southern sky from Australia. 
The main scientific driver is to identify near Earth objects (NEOs) that cast threats to Earth. The three telescopes surveys their own patches of sky, avoiding 10 to 15-degree regions close to the crowded stellar regions on the Galactic plane.
In order to maximize the throughput, the observations were conducted in an un-filtered manner. The exposures were taken in sets of four images, with a time-gap of 10 minutes among images and with typical integration time of 30 seconds. The images were fed to SExtractor \citep{1996A&AS..117..393B} for aperture photometry. Thanks to the high cadence and large patrolling area, the survey is also valuable for time-domain science, which leads to the Catalina Real-time Transient Survey \citep[CRTS;][]{2009ApJ...696..870D}.

The data from the Catalina Schmidt Telescope were obtained in 2005-2013 and can be downloaded from the Catalina Surveys Data Release\footnote{http://nesssi.cacr.caltech.edu/DataRelease/}, with the light curves magnitudes converted to V-band.
In the first data release (CSDR1), \cite{2013ApJ...763...32D} presented $\sim$ 200 million sources with 12$<$V$<$20 magnitudes.
Using this unique data-set, \cite{2014ApJS..213....9D} searched objects in sky regions between declination = -22 degrees and 65 degrees for variables, and
discovered $\sim$ 47,000 periodically varying objects. They further visually inspected the
light curves and presented 4683 detached eclipsing binaries, the largest all-sky eclipsing binary catalog up-to-date, providing a
wealth resource of time-series photometry.

The Catalina Sky Survey light curve of CSSJ114804.3+255132 contains 371 epochs, taken between 2005 and 2013, with a single epoch amounts to a typical exposure time of 30 seconds. The primary and secondary eclipses were observed with $\sim$40 and $\sim$80 epochs, sufficient to sample the eclipses. The photometry data can be retrieved from the Catalina Sky Survey Data Release web-site.
As a starting point, we use the Detached Eclipsing Binary Light curve fitter 
\citep[DEBiL;][]{2005ApJ...628..411D} to analyze the light curve of CSSJ114804.3+255132/SDSSJ114804.35+255132.6.
DEBiL models the binary light curves adopting a simple geometry: it draws an initial guess from approximated analytic formula of detached binaries \citep[see e.g.][]{2003ApJ...585.1038S}, and provides a fit
to several parameters:\\
1. The relative radius ($\frac{R_1}{a}$,$\frac{R_2}{a}$), in units of the semi-major axis $a$.\\
2. The brightness of both stellar components (mag$_1$,mag$_2$).\\
3. The inclination angle ($i$), eccentricity ($e$), and Argument of periastron ($\omega$) of the orbit.\\

DEBiL iteratively fits the light curves by  
minimize the $\chi^2$, adopting the downhill simplex method to find $chi^2$ minima \citep{1965CJ...7...308N} with simulated annealing \citep{1992nrfa.book.....P}. The best-fit DEBiL results and the light curve from CSDR1
are shown in Fig. \ref{fig.lc}. We then pass best-fit parameters from DEBiL to JKTEBOP for further dynamical analysis in Section \ref{sec.dyn}.

\subsection{Spectroscopic follow-ups} \label{sec.spec}

\begin{figure}
  \centering
  \includegraphics[scale=1.0]{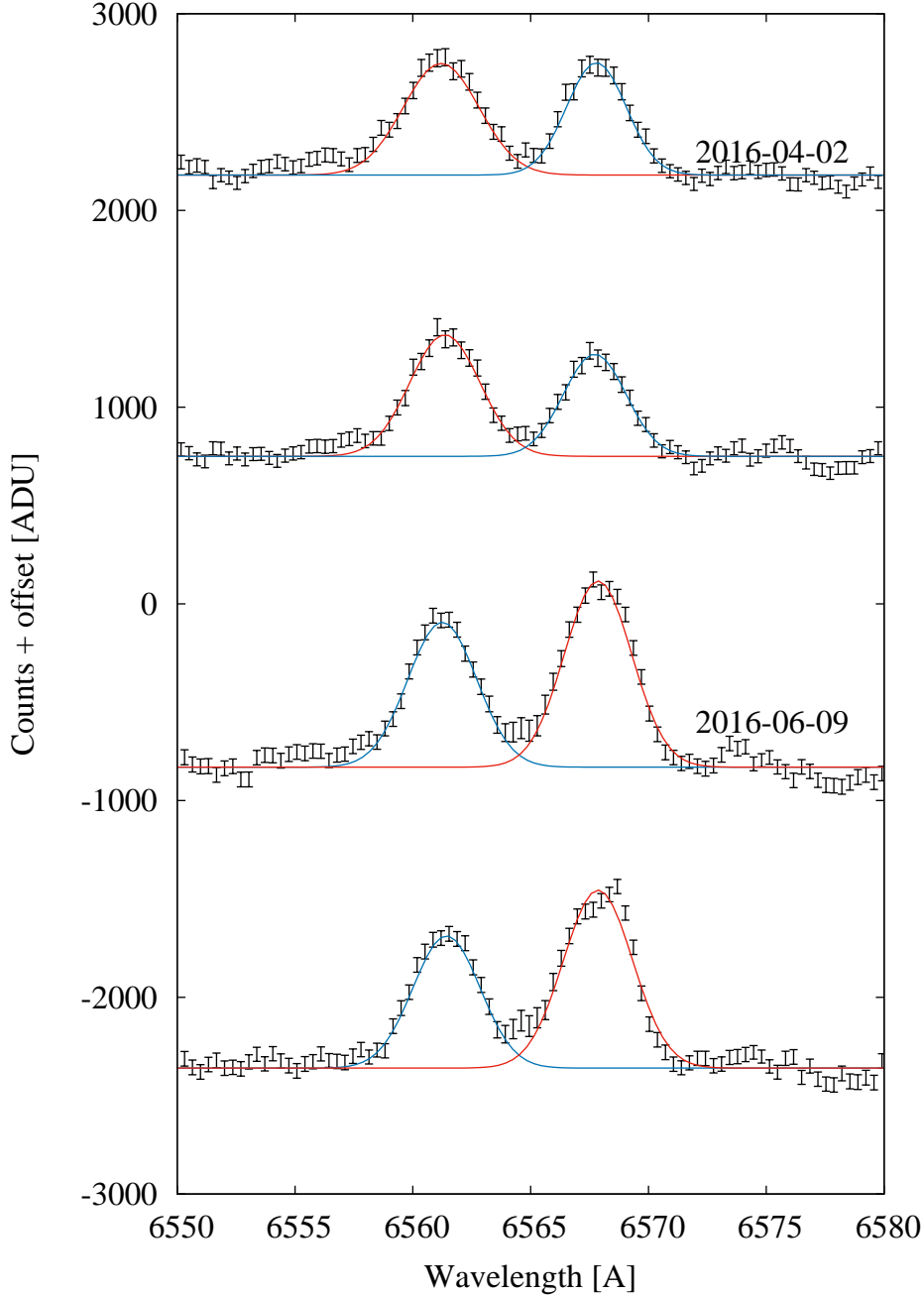}
  \caption{H$\alpha$ emission features from GMOS long-slit observations. From top to bottom are the first and second epochs on 2016-04-02 and 2016-06-09, respectively. The spectra from each epochs are plotted in black. The best-fit Gaussian profiles are plotted in red / blue for the primary / secondary component. Each spectrum is offset arbitrarily in counts for clarity.}
  \label{fig.Halpha}
\end{figure}

\begin{table}
\centering
\caption{Radial velocity from GMOS observations.}
\begin{tabular}[t]{lrr}
  \hline
  \hline
  Epoch & RV$_1$ & RV$_2$ \\
        & [km/s] & [km/s] \\
\hline
2457480.918414 & -73.1 & 228.1 \\
2457480.927963 & -65.8 & 224.9 \\
2457548.774641 & 231.3 & -70.9 \\
2457548.784167 & 231.3 & -62.6 \\
\hline
\hline
\multicolumn{3}{l}{$\dagger$ The estimated radial velocity error is $\pm$ 15 km/s.} 
\end{tabular}
\label{tab.rv}
\end{table}

CSSJ114804.3+255132/SDSSJ114804.35+255132.6 is faint \citep[V=16.85 mag;][]{2014ApJS..213....9D} and its period is very short (P$\sim$0.3 days); to prevent
smooth-out of the radial velocity curve, we need to reach sufficient S/N $\sim$50-100) within 0.1 periods, which is less than an
hour. In addition, to determine the mass and radius of both stellar components, we need to reach sufficient spectral resolution (less
than 0.1nm/pixel). Only 8-m telescope class telescopes are capable of reaching such high S/N and spectral resolution
in the given amount of time. We thus conduct spectroscopic follow-up observations using GMOS \citep{2004PASP..116..425H} on-board Gemini telescope, via fast turnaround program
(Program ID GN-2016-FT16). We
use R831 gratings with 0.5 arc-second slit, enabling a resolution R=4396, with central wavelength at 7000$\AA$ and a coverage of $\sim$2000$\AA$.

The observations were carried out on 2016 April 2nd and June 9th during the expected radial velocity maxima at light curve phase 0.25 and 0.75. In principle we only need one observation to measure the radial velocity maximum, nevertheless we obtain two exposures at each maximum, each with 10-minute integrations to reach high S/N, and the second exposure can serve as a sanity check, ensuring we
obtain consistent radial velocity measurements at the same maximum. Data reduction were carried out using
dedicated IRAF\footnote{http://iraf.noao.edu} GMOS package\footnote{http://www.gemini.edu/sciops/data-and-results/processing-software} (v1.13) in a standard manner. Each spectrum was bias subtracted, flat fielded, sky subtracted,
and wavelength calibrated using CuAr lamp. 

To extract radial velocity information, we make use of the H$\alpha$ emission line at $\lambda\lambda$ 6562.8$\AA$.
The H$\alpha$ emissions from both stellar component are clearly resolved, though with some overlapping (see Fig. \ref{fig.Halpha}). We then
fit the H$\alpha$ line profile using two overlapping Gaussian functions. The best-fit results are shown in Fig. \ref{fig.Halpha}, where
the light contribution for the primary and the secondary component are shown in blue and red, respectively. From the Gaussian fit, we
obtain the radial velocity of each stellar component, as shown in Table \ref{tab.rv}, with estimated error of 15 km/s based on the
spectral resolution delivered by the instrument (i.e. 3.4$\AA$ per pixel).

\section{Analysis} \label{sec.ana}
\subsection{Dynamical analysis} \label{sec.dyn}
\begin{figure*}
  \includegraphics[scale=1.4]{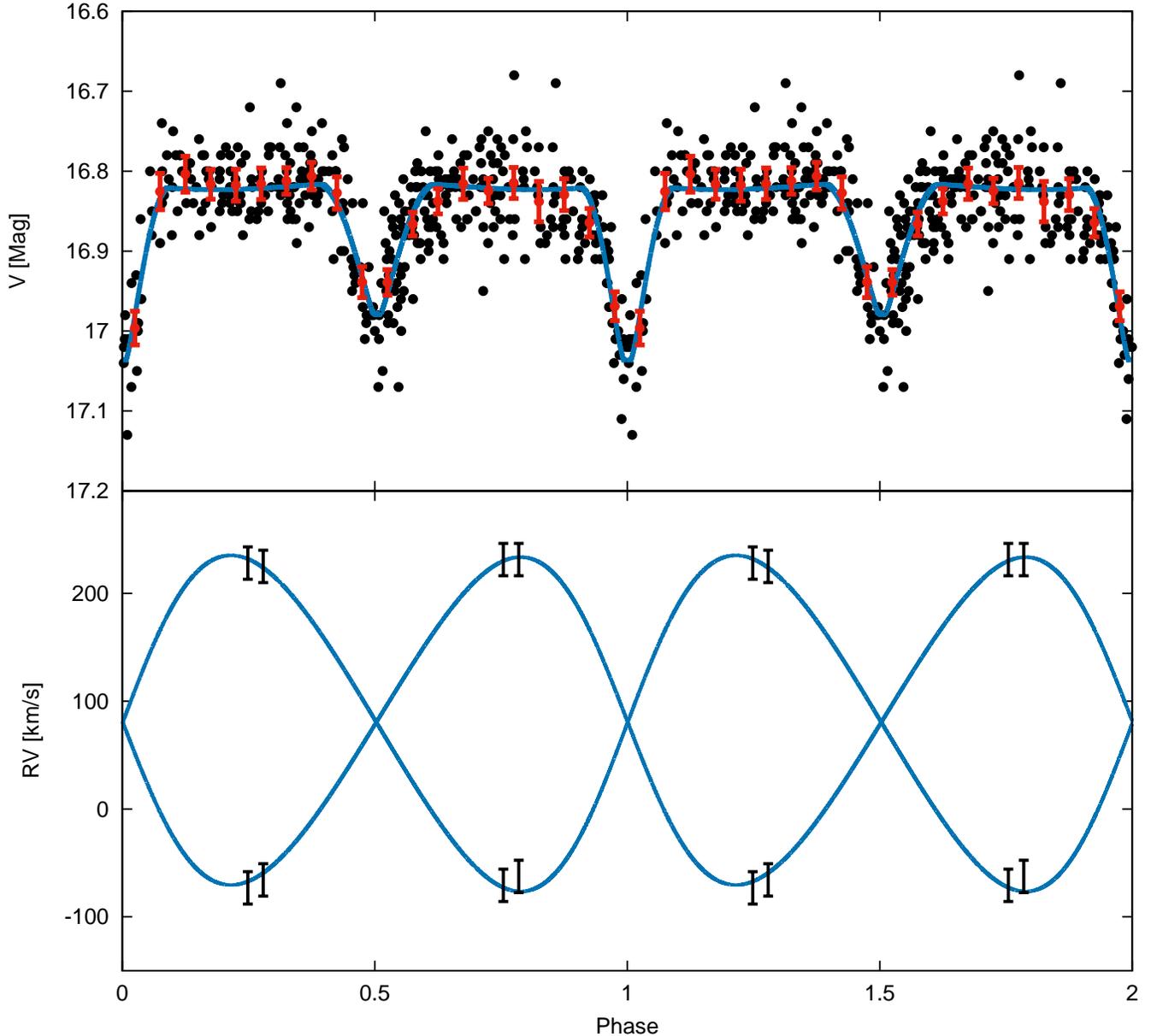}
  \caption{Joint light curve and radial velocity analysis using JKTEBOP. Upper panel (light curve): the single epoch photometry is marked with black circles, the binned data (every 0.05 phase) are shown in red points with error bars, while the best-fit model is shown in blue lines. Lower panel (radial velocity curve): the single epoch radial velocity is marked in black points with error bars, while the best-fit model is shown in blue lines. For illustrational purpose, we plot two full phases to better show the eclipses, with duplicate data shown in phase=1 to phase=2. We note that the eccentricity is poorly constrained and that the eccentric solution shown in this figure does not exclude a (more likely) non-eccentric solution.}
  \label{fig.fit}
\end{figure*}

\begin{figure*}
  \includegraphics[scale=1.8]{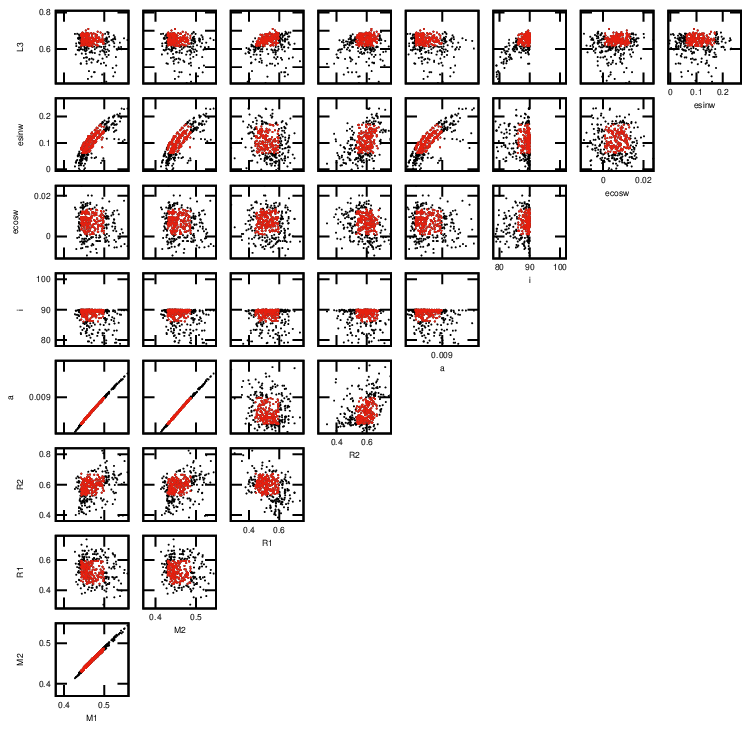}
  \caption{Posterior correlation of the fit parameters assuming an eccentric orbit in JKTEBOP. The red points mark the 68.27 percentile (1$\sigma$) of the MCMC distribution.}
  \label{fig.MC}
\end{figure*}

\begin{figure*}
  \includegraphics[scale=2.5]{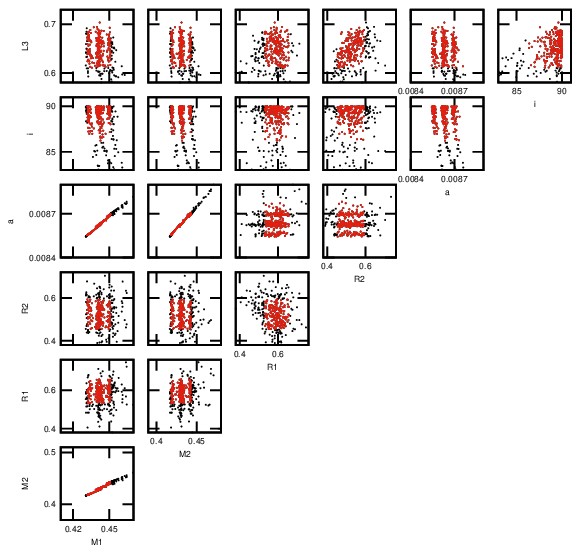}
  \caption{Posterior correlation of the fit parameters assuming an non-eccentric orbit in JKTEBOP. The red points mark the 68.27 percentile (1$\sigma$) of the MCMC distribution.}
  \label{fig.MC2}
\end{figure*}

With the light curve and radial velocity information in hand, we are able to determine the mass and the radius of the system. We use JKTEBOP \citep{2004MNRAS.351.1277S}, a
derivative of the the EBOP code originally written by \cite{1981AJ.....86..102P}, with the capability to jointly fit the light and 
radial velocity curves to determine the mass and radius of each stellar component. The free parameters of the fit are: reference
time of the primary eclipse (t$_0$), radius ratio of the primary to the secondary (R$_2$/R$_1$), radius sum in terms of semi-major axis ((R$_1$+R$_2$)/$a$), inclination angle ($i$), light ratio (L$_2$/L$_1$), orbital
eccentricity ($e$), augment of periastron ($\omega$), third light ratio, radial velocity semi-amplitudes of the two components (K$_1$ and K$_2$). We use the best-fit results from DEBiL in Section \ref{sec.lc} as an initial guess for most of the parameters; for semi-amplitudes K$_1$ and K$_2$, we assume the radial velocity measurements in Section \ref{sec.spec} are approximately the radial
velocity maximum and estimate K$_1$ and K$_2$ are $\sim$ 150 km/s as an initial guess. The JKTEBOP fitting routine 
quickly converges after 22 iterations, indicating the DEBiL results provides a good initial guess. 
The best-fit JKTEBOP results are shown in Fig. \ref{fig.fit} and Table \ref{tab.fit}. The 1-$\sigma$ error in the parameters are estimated using Monte Carlo simulations. We also show the posterior correlations of each parameter in Fig. \ref{fig.MC}. As this is a close binary system, we also perform the fit with eccentricity
fixed to zero. The result of non-eccentric model is also shown in Table \ref{tab.fit} and Fig. \ref{fig.MC2}.

The best-fit eccentric model from JKTEBOP indicates M$_1$=0.47$\pm$0.03 M$_\odot$, M$_2$=0.46$\pm$0.03 M$_\odot$, R$_1$=0.52$\pm$0.08 R$_\odot$, 
and R$_2$=0.60$\pm$0.08 R$_\odot$. While the primary component is in good agreement with the empirical mass-radius 
relation of \cite{2009A&A...505..205D}, the secondary component's radius appears to be inflated (by 41\%), 
which has been seen in short-period systems and can be attributed to the tidal locking of the close companion 
star \citep{2011ApJ...728...48K}.
On the other hand, for a non-eccentric model, JKTEBOP returns
M$_1$=0.44$\pm$0.01 M$_\odot$, M$_2$=0.43$\pm$0.01 M$_\odot$, R$_1$=0.59$\pm$0.06 R$_\odot$, 
and R$_2$=0.52$\pm$0.07 R$_\odot$. While the secondary component is in good agreement with the empirical mass-radius 
relation of \cite{2009A&A...505..205D}, the primary component's radius appears to be inflated (by 44\%). We suggest to use the difference between the eccentric and non-eccentric fit as an estimate of the systematic error, which at the level of 0.01 M$_\odot$ for both primary and secondary masses, and at the 0.07 and 0.08 R$_\odot$ level for the primary and secondary radius, respectively.
From the posterior correlation plots, we also found degeneracy between the masses (M$_1$ and M$_2$), orbital distance ($a$), and eccentricity (in terms of esin$\omega$), as shown in Fig. \ref{fig.MC}. We note that the eccentricity is expected to be zero physically, and the poor phase coverage of the raidal velocities means that the eccentricity is not well-constrained, as can be seen from the difference in the fits.

\begin{figure*}
  \includegraphics[scale=0.6]{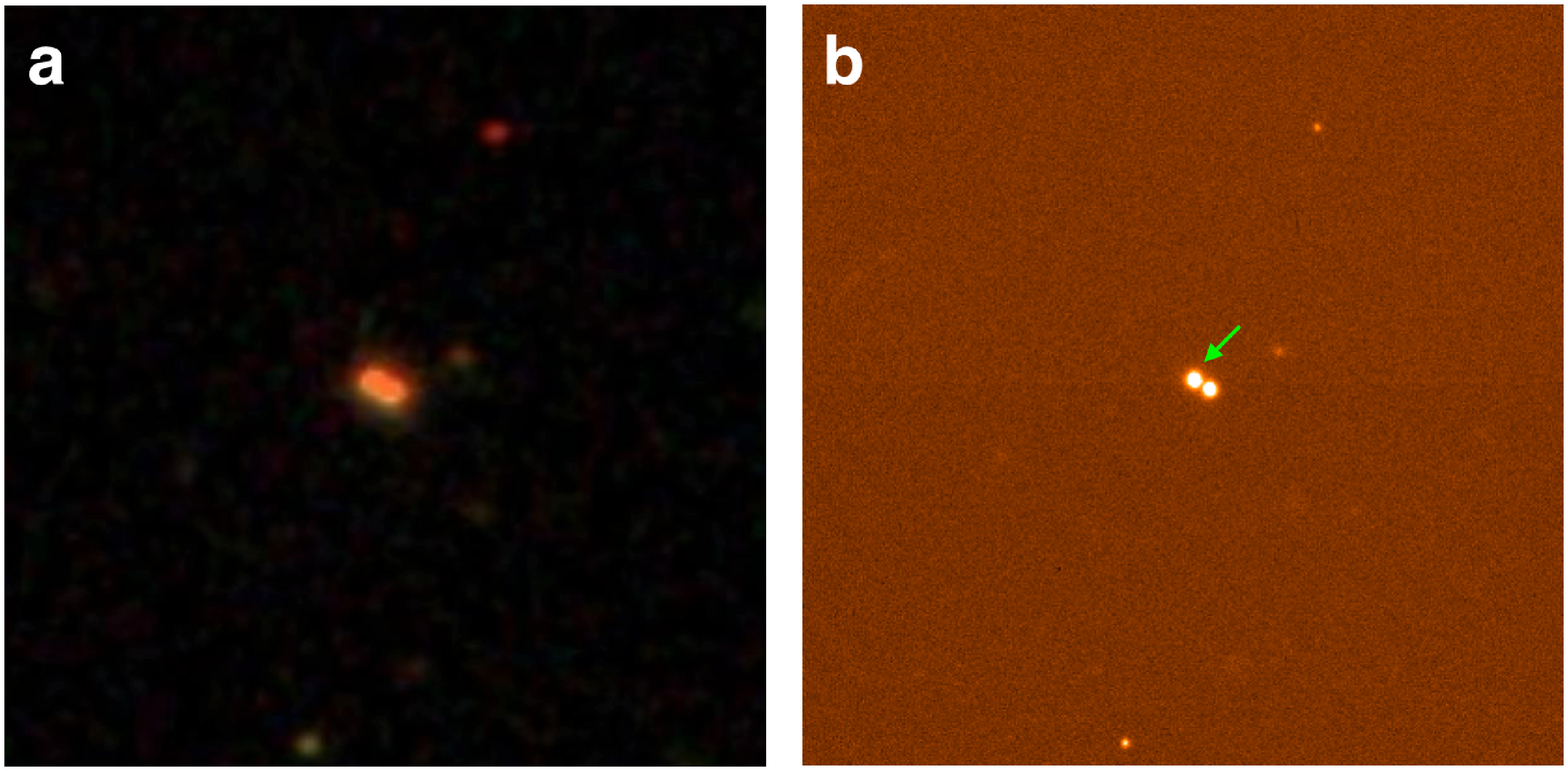}
  \caption{Zoom-in imaging of the binary system. North is up, with east to the left. The size of the images are 1x1 arcmins. (a) SDSS finding chart imaging, composited
    of g, r, and i-band filters. The binary system and the third object are partially resolvable. (b) Gemini acquisition imaging in r-band. The binary system is indicated
    by the green arrow, where the third object (at a separation of 1.5 arcseconds) is well resolved. The Gemini spectra were carried out with a position
  angle of 90 degrees east of north, hence were not contaminated by the third object.}
  \label{fig.vlms1}
\end{figure*}

The best-fit third light is relative high (66\%), which can
explain the shallow eclipse depth for a flat-bottomed eclipse.
Such high third light contribution indicates that we may see the third object with better spatial resolution.
  We exploit the acquisition imaging of the Gemini spectroscopic observation (see Fig. \ref{fig.vlms1}) and found out a third object $\sim$1.5 arcseconds
  away from the binary system. Such small separation is not resolvable from the poor spatial resolution of CSS, and partially resolvable
  from the SDSS imaging (see Fig. \ref{fig.vlms1}). The SDSS spectroscopic observations were carried out with a
  fibre diameter of 3 arcseconds and centered on the binary system rather than on the third object, which avoid contamination from this third object in the spectra.
  The Gemini follow-up spectra were obtained by putting a 0.5-arcsecond slit centered on the binary system with a positiong angle of 90 degrees east of north,
  so they were also not contamined by the third object.

\begin{figure*}
  \includegraphics[width=\columnwidth]{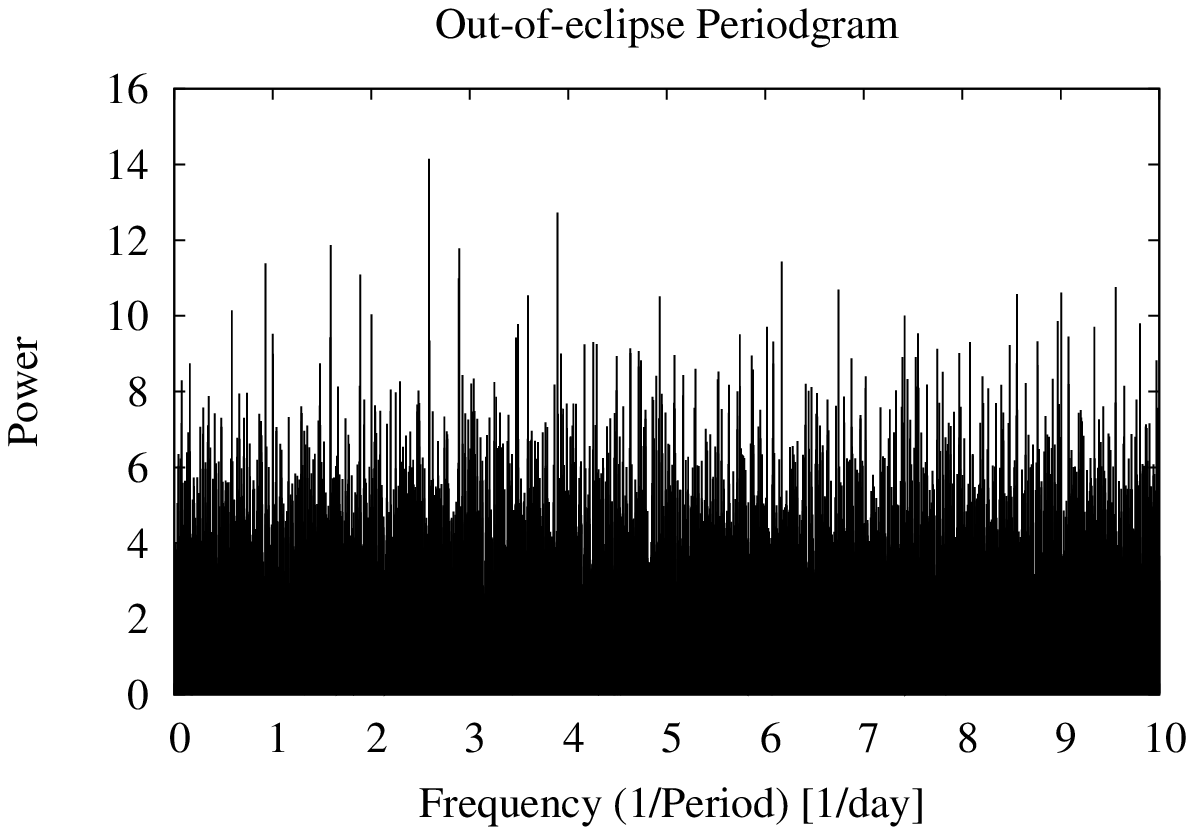}
  \caption{Periodogram of out-of-eclipse light curves, using Lomb-Scargle algorithm. We intend to detect periodic
  photometric modulations as a sign of stellar spot, but there is no clear evidence from periodogram.}
  \label{fig.spot}
\end{figure*}

Another important point to take into consideration
when modeling eclipsing binaries is the effect of stellar spots. To investigate the presence of stellar spots, we thus
use the out-of-eclipse light curves to search for periodic modulations as a sign of stellar spots. However,
we did not find significant variations or periodicity using Lomb-Scargle algorithm, as shown in Fig. \ref{fig.spot}.
Since spots are not evident, a spot analysis is not performed.

\subsection{Temperature}

\begin{figure*}
  \includegraphics[scale=0.72]{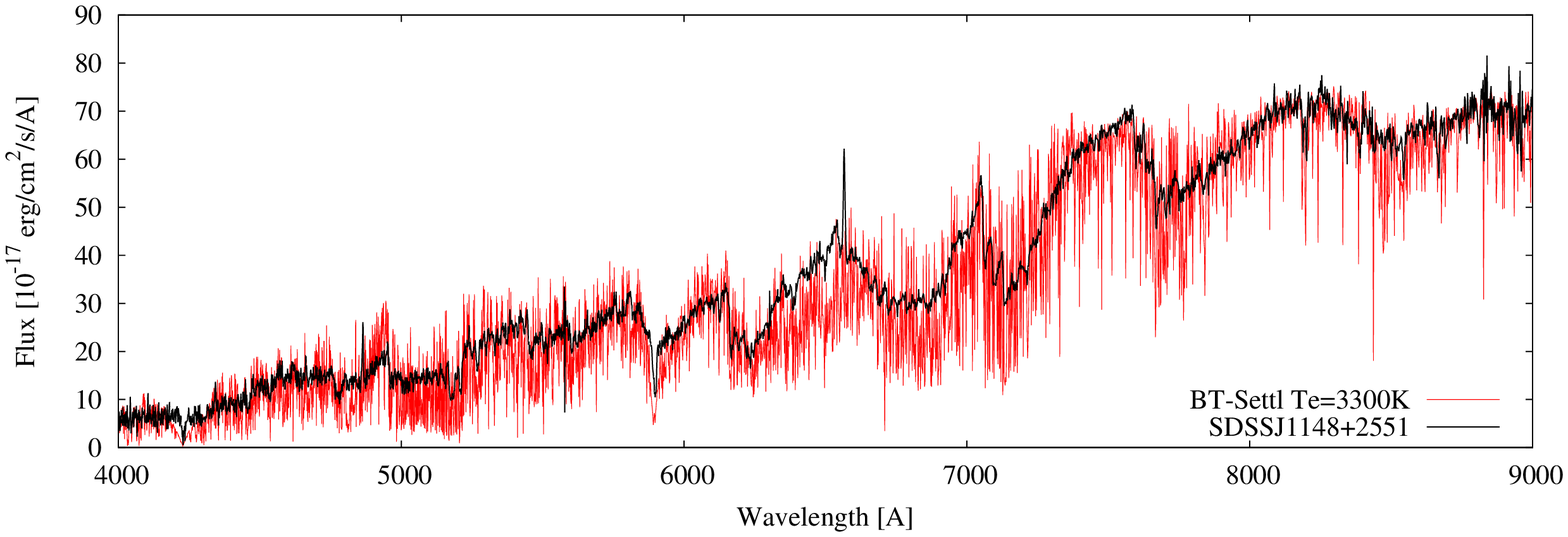}
  \caption{Single shot spectra from SDSS (marked in black) with the best-fit BT-Settl model spectrum (marked in red).}
  \label{fig.spec}
\end{figure*}

To estimate the temperature, we make use of the SDSS single shot spectrum, which covers the entire optical wavelength from 4000-9000 $\AA$. SDSS spectrum has relative low resolution (R$\sim$1800), hence can be regarded as a spectrum from a 
single star, instead of the binary sources. To derive the temperature of the binary systems, we conduct template spectrum fitting in a grid manner, similar to the approach of \cite{2015ApJ...804...64M}. We then fit a set of template M dwarf spectrum from 
BT-Settl model \citep{2012EAS....57....3A}, assuming 
log$g$=5 (consistent with the mass and the radius from dynamical modeling in the previous section) and solar 
metallicity. We use a grid of temperatures, ranging from 2500 to 3500 K, with a step size of 100 K. 
The best-fit BT-Settl model indicates a temperature of 3300K (see Fig. \ref{fig.spec}. To dissect the 
temperature of each stellar component, we follow the procedure of \cite{2015MNRAS.451.2263Z} and make use of the light ratio ($\frac{L2}{L1}=0.28$) and the radius ratio estimated in 
previous section. Assuming black body radiation, $\frac{L2}{L1}\propto\frac{R_2^2T_2^4}{R_1^2T_1^4}$, we derive a temperature of T$_1$=3560K and T$_2$=3040K, respectively, 
with an estimated error of 100K. 

\begin{table}
\centering
\caption{Best-fit parameters for CSSJ114804.3+255132/SDSSJ114804.35+255132.6}
\begin{tabular}[t]{lrr}
  \hline
  \hline
  Parameter & Free ecc. model & Non-ecc. model\\
  \hline
  \multicolumn{3}{c}{\underline{\it Modeled parameters}} \\
 
t$_0$ [HJD] & 2454096.8987$\pm$0.0007 & 2454096.8992$\pm$0.0007\\
(R$_1$+R$_2$)/$a$ & 0.59$\pm$0.04 & 0.60$\pm$0.04\\
R$_2$/R$_1$ & 1.16$\pm$0.24 & 0.88$\pm$0.17\\
$i$ [deg] & 89.9$\pm$3.92 & 89.94$\pm$3.48\\
e cos$\omega$ & 0.007$\pm$0.006 & Fixed to 0 \\
e sin$\omega$ & 0.12$\pm$0.05 & Fixed to 0\\
Third light & 0.66$\pm$0.05 & 0.66$\pm$0.05\\ 
K$_1$ [km/s] & 152.06$\pm$3.83 & 147.82$\pm$1.69\\
K$_2$ [km/s] & 155.89$\pm$3.59 & 151.61$\pm$0.62\\
T$_1$ [K] & 3560$\pm$100 & 3560$\pm$100\\
T$_2$ [K] & 3040$\pm$100 & 3040$\pm$100\\
\hline
\hline
\multicolumn{3}{c}{\underline{\it Derived parameters}} \\
\hline
M$_1$ [M$_\odot$] & 0.47$\pm$0.03 & 0.44$\pm$0.01 \\
M$_2$ [M$_\odot$] & 0.46$\pm$0.03 & 0.43$\pm$0.01 \\
R$_1$ [R$_\odot$] & 0.52$\pm$0.08 & 0.59$\pm$0.06 \\
R$_2$ [R$_\odot$] & 0.60$\pm$0.08 & 0.52$\pm$0.07 \\
a [AU] & 0.0093$\pm$0.0003 & 0.0086$\pm$0.0001 \\
L2/L1 & 0.71$\pm$0.09 & 0.63$\pm$0.27\\
\hline
\hline
\end{tabular}
\label{tab.fit}
\end{table}

\section{Summary and prospects} \label{sec.sum}
We present a preliminary analysis of the double-lined M dwarf eclipsing binary system
CSSJ114804.3+255132 / SDSSJ114804.35+255132.6, discovered by cross-matching eclipsing binaries charted
by Catalina Sky Survey and spectroscopic confirmed M dwarfs in SDSS. We obtained follow-up medium resolution
spectra using GMOS on-board Gemini telescope, enabling us to disentangle the emission lines from both stellar
components, and providing us the radial velocity information during radial velocity curve maxima. Joint analysis
of the light and the radial velocity curves using JKTEBOP code indicates that, under the context an eccentric model,
the secondary's radius is large (by 41\%) compared to the empirical mass-radius relation of low mass stars, while the primary shows agreement with such relations. On the other hand, if we assume a non-eccentric model, the primary's radius is inflated (by 44\%), while the secondary shows
  agreement with the empirical mass-radius relation. The inflated stellar component might originate from tidal locking effect exerted by the
close companion, inducing enhanced stellar activity and inhibited convection, commonly seen in short period
binaries. 
Future high resolution spectroscopic observations will help pin down the uncertainties in the fundamental
parameters of this system. Multi-epoch spectra, focusing on H$\alpha$ or other indicator of stellar activities,
will also shed light on the strength of stellar activity, to test the tidal-locking induced inflation scenario.
The system would also significantly benefit from higher spatial resolution photometry. This could likely
be carried out with smaller telescopes, given the depth of the eclipse (in the absence of 3rd light).

Besides SDSS, currently there is also another large area spectroscopy survey LAMOST, using the 6-m LAMOST telescope
fed with multi-object fibres to characterize stellar objects and galaxies in the northern sky. With sky coverage
similar to SDSS, we anticipate there will be several M dwarf eclipsing binaries revealed by LAMOST. In addition,
the planned all-sky spectroscopic survey 4MOST will also enable similar search for spectroscopic M dwarfs
in the southern sky. As most of the large area spectroscopic surveys are already observing very faint objects and
discovered many M dwarfs, the key breakthrough will be made by future all-sky surveys with larger telescopes, e.g. LSST.
This is because the current all-sky time-domain studies hardly go beyond V=18th magnitude or fainter, where the vast
majority of M dwarfs will be dimmer than 19 magnitudes. However, it is going to be challenging to obtain radial velocity
measurements for 19th magnitude M dwarf binaries from LSST.

\acknowledgments
We are indebted to the anonymous referee, whose comments greatly improved the manuscript.

The CSS survey is funded by the National Aeronautics and Space Administration under Grant No. NNG05GF22G issued through the Science Mission Directorate Near-Earth Objects Observations Program. The CRTS survey is supported by the U.S.-National Science Foundation under grants AST-0909182.

Funding for the Sloan Digital Sky Survey IV has been provided by
the Alfred P. Sloan Foundation, the U.S. Department of Energy Office of
Science, and the Participating Institutions. SDSS-IV acknowledges
support and resources from the Center for High-Performance Computing at
the University of Utah. The SDSS web site is www.sdss.org.

SDSS-IV is managed by the Astrophysical Research Consortium for the 
Participating Institutions of the SDSS Collaboration including the 
Brazilian Participation Group, the Carnegie Institution for Science, 
Carnegie Mellon University, the Chilean Participation Group, the French Participation Group, Harvard-Smithsonian Center for Astrophysics, 
Instituto de Astrof\'isica de Canarias, The Johns Hopkins University, 
Kavli Institute for the Physics and Mathematics of the Universe (IPMU) / 
University of Tokyo, Lawrence Berkeley National Laboratory, 
Leibniz Institut f\"ur Astrophysik Potsdam (AIP),  
Max-Planck-Institut f\"ur Astronomie (MPIA Heidelberg), 
Max-Planck-Institut f\"ur Astrophysik (MPA Garching), 
Max-Planck-Institut f\"ur Extraterrestrische Physik (MPE), 
National Astronomical Observatory of China, New Mexico State University, 
New York University, University of Notre Dame, 
Observat\'ario Nacional / MCTI, The Ohio State University, 
Pennsylvania State University, Shanghai Astronomical Observatory, 
United Kingdom Participation Group,
Universidad Nacional Aut\'onoma de M\'exico, University of Arizona, 
University of Colorado Boulder, University of Oxford, University of Portsmouth, 
University of Utah, University of Virginia, University of Washington, University of Wisconsin, 
Vanderbilt University, and Yale University.

Based on observations obtained at the Gemini Observatory and processed using the Gemini IRAF package,
which is operated by the Association of Universities for Research in Astronomy, Inc., under a cooperative
agreement with the NSF on behalf of the Gemini partnership: the National Science Foundation (United States),
the National Research Council (Canada), CONICYT (Chile), Ministerio de Ciencia,
Tecnolog\'{i}a e Innovaci\'{o}n Productiva (Argentina), and Minist\'{e}rio da Ci\^{e}ncia, Tecnologia e Inova\c{c}\~{a}o (Brazil).

The authors wish to recognize and acknowledge the very significant cultural role and reverence that the summit of Maunakea
has always had within the indigenous Hawaiian community.  We are most fortunate to have the opportunity to conduct observations from this mountain.


\end{document}